# Results of the experiment on investigation of Germanium-76 double beta decay

A.M. Bakalyarov, A.Ya. Balysh, S.T. Belyaev, V.I. Lebedev, S.V. Zhukov

## Abstract

On the termination of the Heidelberg-Moscow double-beta decay of Germanium-76 experiment, the analysis of the obtained data is presented. The value of the half-life for two-neutrino double beta decay (2β2ν mode) and limitation of the half-life for neutrinoless double beta decay (2β0ν mode) are given. The results were presented at NANP2003, Dubna, June, 24.

## Experimental setup

High purity germanium crystals, enriched by Germanium-76 isotope up to 86% are used as the main detecting elements.  Five coaxial detectors with the total weight of 11,5 kg (125 moles in the active volume of detectors) are used.

Each detector is located in a separate cryostat made of electrolytic copper with low content of radioactive impurities. The quantity of other designed materials (iron, bronze, light material insulators) is minimized in order to reduce the feasible radioactive impurities contribution to the total background of the detectors.

The detectors are located in two separate shielded boxes. One of them, 270 mm thick is made of electrolytic copper (detector #4), the other consists of two layers of lead – inner - 100mm of high purity LCD2-grade lead and outer – 200mm of low background Boliden lead (detectors ## 1,2,3,5). Each setup is coated with stainless steal casing. Non-radioactive pure nitrogen was blown through casings to reduce radon emanation contribution. To reduce neutron background the casing with detectors ##1,2,3,5 was coated with borated polyethylene and two anticoincidence plates of plastic scintillator were located over the casing in order to reduce muon component.

The setup is located in Gran Sasso underground laboratory, Italy. The depth of 3500 m of water equivalent of the lab reduces influence of cosmic rays on background conditions of the experiment.

The electronics and the system of collecting data allow to record each event – the number (or numbers) of acted detector, amplitude and pulse shape, and anticoincidence veto. Detailed descriptions of the setup, stability control system and calibration procedure are in [1].

Experimental data obtained within the period since November 1995 till August 2001 (200 days after the last detector started working in the setup.) are used in this work .In terms of active volume statistical significance for all the five detectors is equal to ~ 50,5 kg*year.

## Background model

For correct interpretation of data obtained throughout the experiment, it is necessary to construct a model, which is the most suitable for the background conditions of the experiment.  By varying the value and location of the background sources one can reach the best agreement between the model and experimental spectrum. The uniform distribution of radioactive impurities in each material was the only a priori condition.

Response functions of each detector to products of decay of all isotopes of Uranium and Thorium radioactive chains were computed with the help of Monte Carlo method (GEANT3-2.1). Gamma quanta, electrons and bremsstrahlung, accompanying beta decay were taken into account in the simulation. The response functions for other background sources such as K-40, Co-60, Cs-137, Cs-134, Bi-207, Sb-125, Mn-54, Zn-65, were reliably identified by characteristic lines. In addition, the bremsstrahlung spectra of Bi-210, as well as spectra of Co-60, Ga-68, Ge-77, Sr+Yt-90 and some other isotopes, presented, as small impurities in Germanium crystals were included in computation. These impurities can contribute in total background, though without creating characteristic peaks, which identify them. The contribution of neutrons to background spectrum was computed as well.

Consideration of surface contamination was of importance, as well. α-peaks of Po-210, which are vividly seen in spectra of detectors ##4,5 stand for that possibility. The high energetic range of spectrum is well described by α-particles of Th and U radioactive chains. It is shown on figure 1.

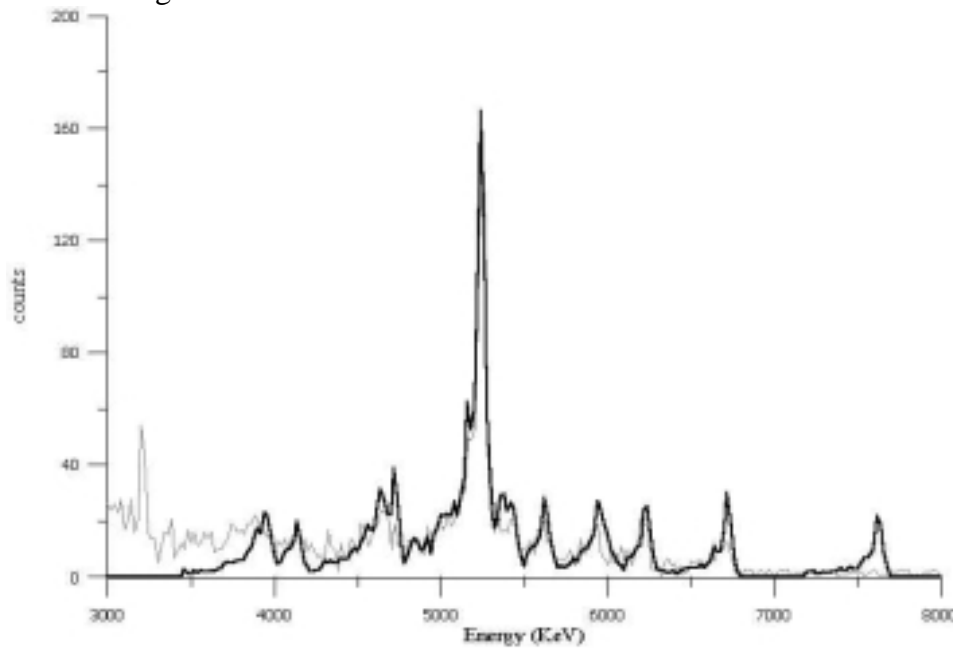

Fig.1 Sum of spectra of all five detectors in high energetic range, summarized in intervals of 20 keV. Pale line is experimental spectrum; thick line is simulated spectrum of α-particles of Thorium and Uranium radioactive chains, which are in equilibrium with decay products. The shape of spectrum was computed out of shape of experimental α-spectrum of Po-210

The peaks are shifted to the left on 80 – 100 keV and have asymmetric shapes. All this shows that α-sources are located on the surface of the crystals. The spectrum calculated with the assumption, that α-particles are passing through the dead zone in the detector well is in good agreement with observation. It is quite evident that products of Uranium and Thorium chains are close to surface of crystals, and that was also taken into account when constructing the background model.

<u>A few technical details</u>

Each time a supposed source of radioactive contamination was uniformly distributed in the volume (surface) of construction materials of the detectors and shielding. Contribution of each isotope was computed separately. Only information about quantity of detected events in well-identified peaks was used. Search for minimum of function $F_{min}$ for all the five detectors simultaneously was the criterion for correct description of spectrum, where

$$F_{min} = \sum_{i=1}^{i} \sum_{j=1}^{j} (x_{i,j}(sim) - x_{i,j}(exp))^2 / \sigma^2_{i,j}(exp) \qquad (1),$$

Here - i is the number of the detector,
- j is the number of the peak,
- x (sim) and x (exp) are areas of experimental and simulated peaks,
- σ (exp) is statistical error in area of experimental peak.

The square of computed peak - x (sim) is determined as a sum of all materials contributions

$$x(sim) = \sum_{n=1}^{n} x_n \qquad (2),$$

- n is a number of simulated parameters (materials).

This method allows to locate isotopes, which have several peaks in spectrum (at least two peaks) with high accuracy. To locate isotopes with one peak in spectrum (Cs-137, K-40) or without peaks (Pb-210), we used additional information, which we got from preliminary experiments (before 1995), when detectors worked when incompletely assembled or inside individual shielding. Particularly when using this method we found and located Cesium spot on inner surface of lead shielding which explains relatively high background of Cesium in the detector #5.

After that, contributions of separate isotopes computed before were considered as constants, but isotopes, which presence in Germanium itself is possible and also 2β2ν mode were varied separately for each detector. Bremsstrahlung of Bi-210 presenting in lead shielding varied for detectors ##1,2,3,5 jointly.

The procedure of fitting was repeated many times for various low boundaries from which fitting started (low boundary varied from 200 to 1800keV).

As a result of fitting procedure mentioned above, a constrained range of the most probable localization parameters for background sources was found. Uranium and Thorium radioactive chains as well as Cobalt and Potassium in different combinations and 2β2ν mode varied simultaneously. As a result of these fittings the half-life for 2β2ν mode of Germanium-76 was calculated

$$T_{1/2}(2\beta2\nu) = (1.78 \pm 0.01 \, [stat] \, ^{+0.07}_{-0.09} \, [sist]) * 10^{21} \, y \quad (68\% \, C.L.) \qquad (3)$$

The systematic error is mostly determined by inaccurate localization of background sources and also includes uncertainty in determination of active volume and degree of enrichment of Germanium. Figure 2 illustrates half-life for 2β2ν mode of Germanium-76.

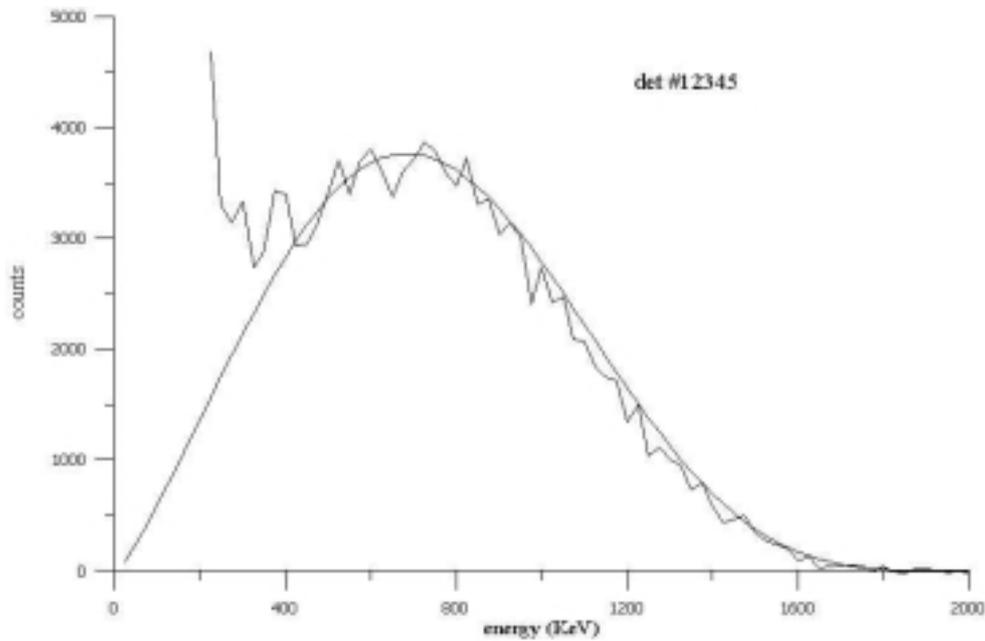

Fig. 2 Summarized spectrum of five detectors, integrated in 25 keV Broken line is a difference between experimental data and simulated background; smooth line is a computed value of half-life for two-neutrino mode.

Analysis of the spectrum in the range of 2β0ν mode

Suggested model allows to describe the background in the neutrinoless beta decay region (see fig. 3).

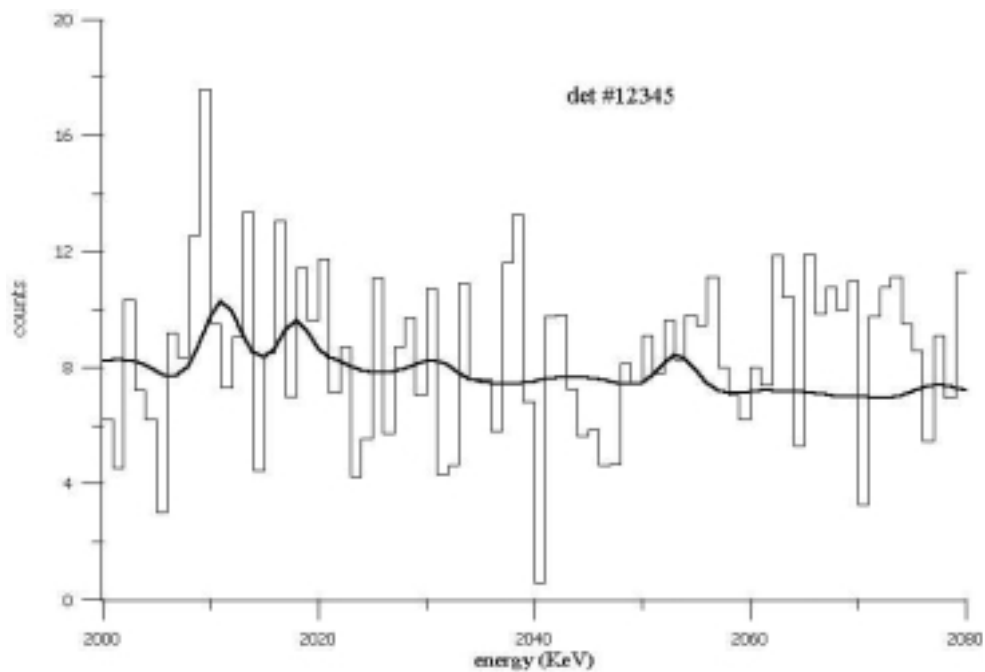

Fig. 3

The histogram is experimental data; smooth line is a computed background.

The model describes around 90% of experimental data in 2000 – 2080 keV region (in 2000 – 2060 keV region it describes ~ 98%). The important circumstance is that the background is practically constant in this region under investigation (summarized quantity of events in three peaks of Bi-214 is less than 5% from total background).

## Reliability of experimental data

Estimation of reliability of the experimental data was carried out in parallel with the calculation described above. The aim was to find out short-live components and test the apparatus stability. To make sure that all statistics corresponds to Poisson distribution and apparatus works correctly, the experimental data was randomly divided in several samples, and their spectra were compared. In particular, this analysis showed significant variations in time of Bi-214 and Pb-214 peaks, which stands for noticeable contribution to the background from Rn-226 inside the shielding. At the same time we found a number of peaks, which could not be identified; their intensity changed arbitrarily for different detectors from sample to sample. They were better seen in the spectra of the detector #4. More careful analysis showed that those peaks appeared in the runs where there were pulses under discriminator threshold ("under_level"or underthreshold) and did not appear if there were no such pulses. In figure 4 one of such peaks is shown.

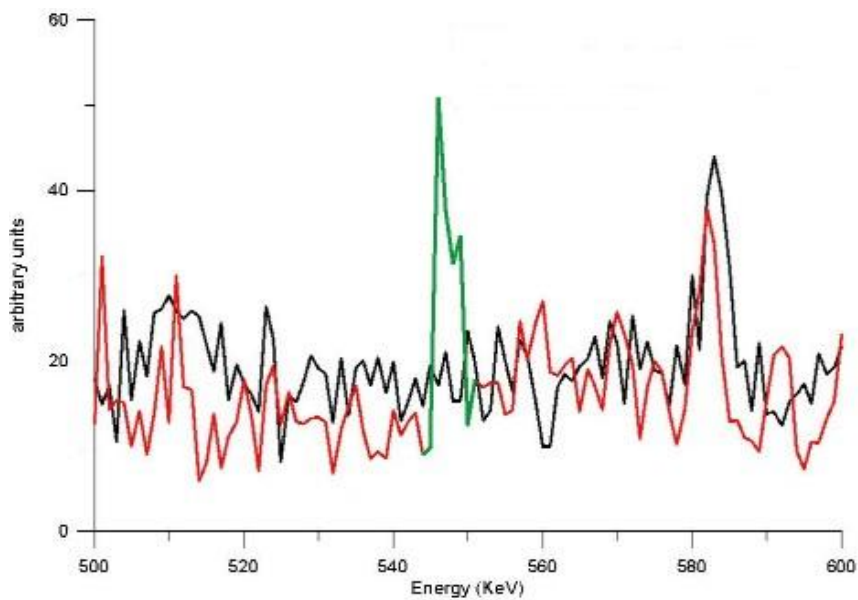

Fig. 4
Detector # 4.
Black line corresponds to runs without underthreshold pulses; red line corresponds to runs with underthreshold pulses (green line is unidentified peak). Spectra are normalized by time.

Figure 5 shows characteristic amplitude spectra of one of the detectors (#5) in the range of low threshold of discriminator are performed. The spectra were not normalized and their sum corresponds to full time of measurements with this detector. Pulses which occur lower, than the discriminator threshold (dashed line) are vividly seen. The runs with such pulses were tagged as "under_level".

Fig. 5
Detector # 5.
Amplitude spectra in the low energy range.
Dashed line - corresponds to runs with underthreshold pulses ("under_level"); red line corresponds to runs without underthreshold pulses.

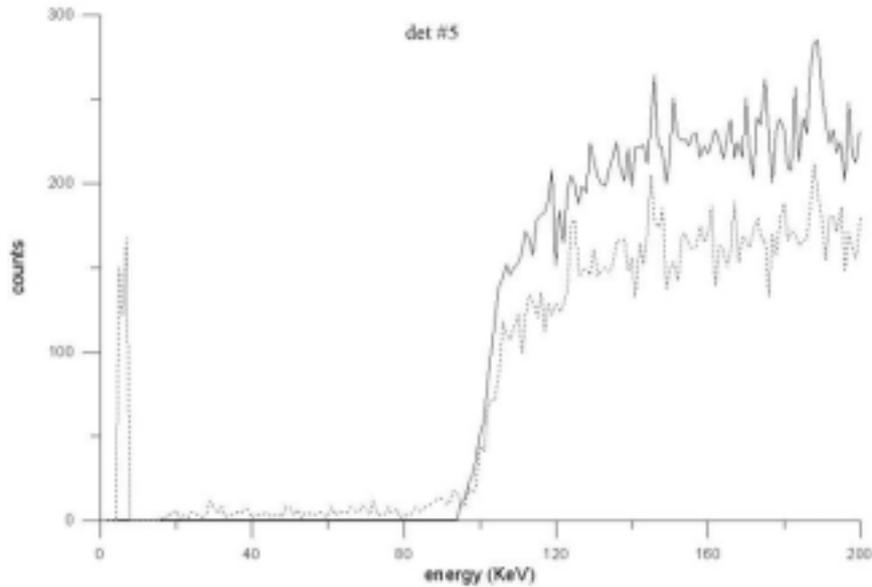

The origin of the under threshold pulses and false peaks cannot be disclosed without special control experiments. As a matter of fact, since underthreshold pulses did not occur in all runs, their quantity is quite small (1 – 3 per 100 events, typical for run of separate detector) and they were out of main energy range, we could not take them into consideration. But the fact that unidentified peaks appear (with the lower probability also in other detectors and in different regions) required more precise investigation of experimental spectra, in order to analyze the influence of those events on evaluation of physical result

For this purpose, the set of the experimental data was divided into two subsets – with underthreshold events and without them. After that for each of the two subsets (normalized by total statistics) the procedure of the calculation of two-neutrino mode for the whole statistics described above was used.

The values of 2ν- mode for the two subsets, normalized by the time of measurement are shown in figure 6. Two curves are in good accordance and the numerical difference might be taken into account in systematic error.

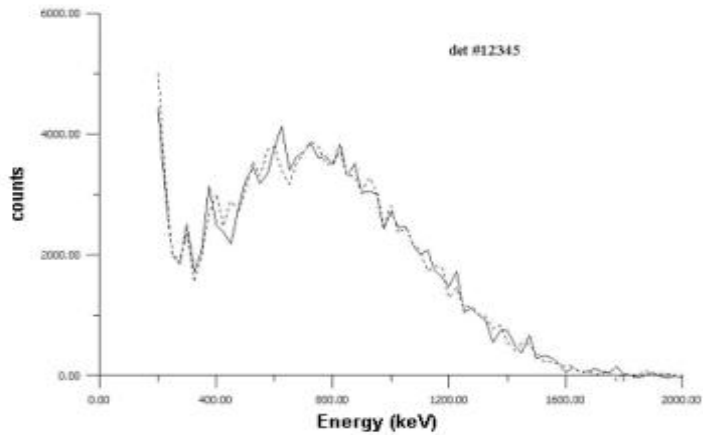

Fig. 6

Comparison of computed values of 2ν-mode for two sets; without "under_level" pulses – solid line; "under_level" – dashed line.

Neutrinoless mode with its characteristic peak requires more specific analysis. All isotopes, which could get the peak in energy range 2036- 2042 keV, were investigated with great accuracy. It was shown that none of them could be responsible for appearance of peak in that energy range. It was decided to compare spectra with underthreshold events with spectra without such events. Statistical significance of those two sets is comparable.

Since false peaks do not appear in all runs with underthreshold events, a general view of spectra is close enough. But in some energy ranges the difference is rather visible. Figure 7 illustrates comparison between two discussed sets for detector #4 (it is located in individual shielding).

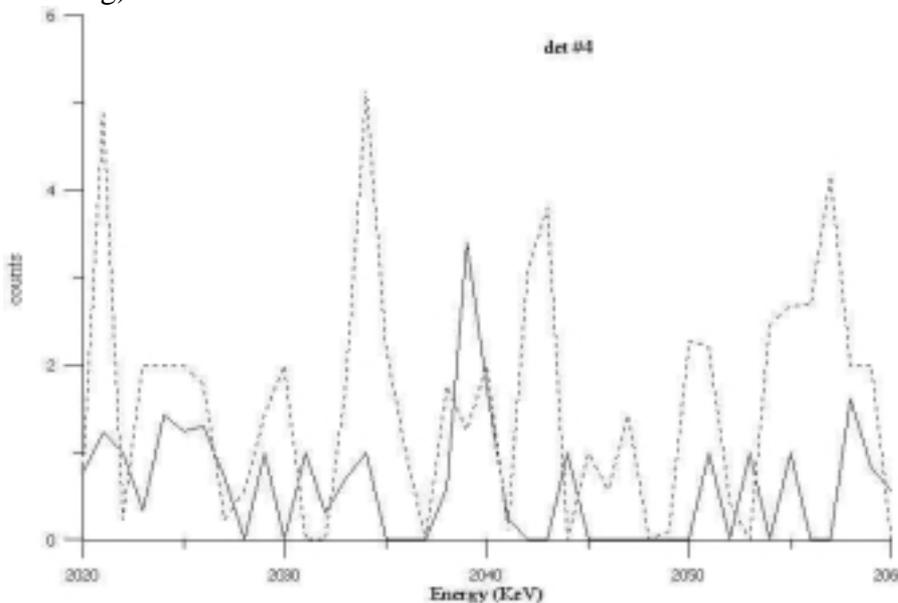

Fig. 7
Black line corresponds to runs without under threshold events, dashed line corresponds to runs with under threshold events

Analogues comparison of spectra for neutrinoless double beta decay region from detectors ##1,2,3,5 seems to be of more interest.

Two summarized in 2039 keV range spectra of detectors ##1,2,3,5 with and without under threshold events are shown in figure 8

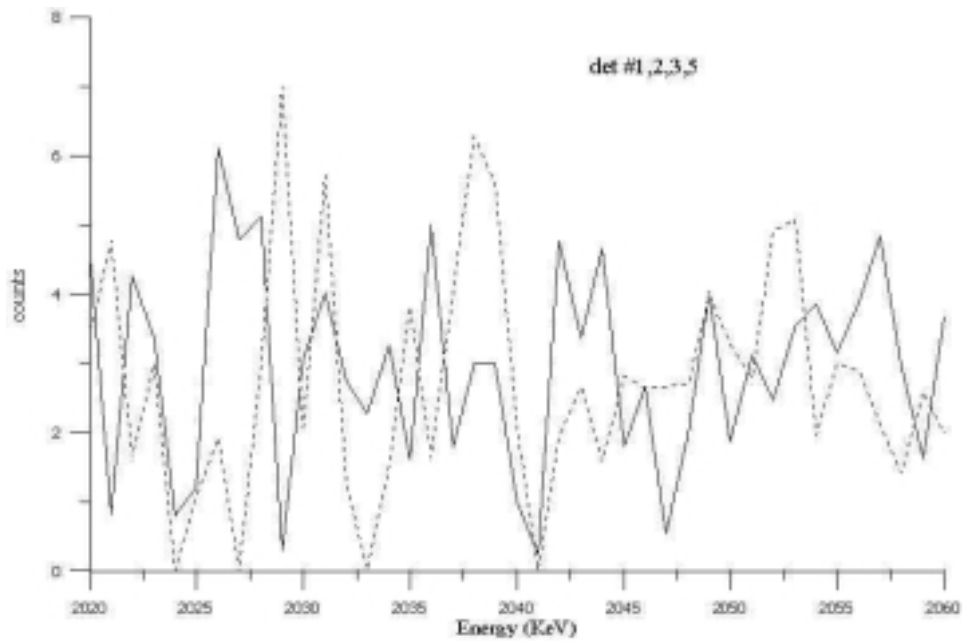

Fig. 8
Black line corresponds to runs without under threshold events; dashed line corresponds to runs with under threshold events. The data are not normalized; the time of collecting statistic for assembling of four detectors is approximately equal (47% for "under_level" runs and 53% without those runs).

When looking at this figure one can doubt that peak of 2039 keV has physical nature. Treatment of this part of the spectrum might be carried out by means of "Bayesian method", used in work [2], where the experimental spectrum in 19 KeV interval with the center at 2039 KeV was considered as a sum of constant background and one gaussian with determined energy resolution. Algorithm permits to get probability density function of Gaussian admission in the whole spectrum. The outcome of this procedure with the same energy window as in [2] shows that the peak appears in "under_level" runs and it does not appear where there are no underthreshold events.

It is necessary to mention that this peak is obtained only in the detectors ##3,5. Spectrum of total statistics of detectors ##1,2 and runs of detectors ##3,5 without "under_level runs are presented in figure 9. Statistical significance of this set corresponds to 70% of total statistics. This figure clearly testifies against the presence of the peak. The treatment of the spectrum by Bayesian algorithm suggested in [2] just confirms this result.

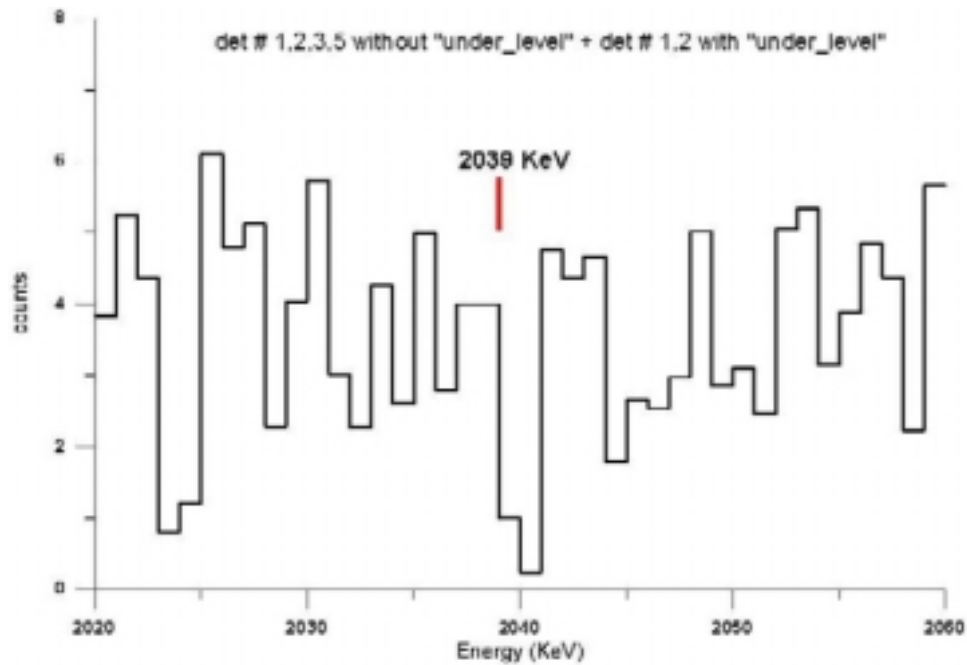

Fig 9
The whole statistics of four assembled detectors without "uder_level" events of detectors ##3,5

It is interesting to compare the integrals of Gaussian peaks, computed with the "Bayesian algorithm" (best value) for three sets.
1. 14,3 – detectors ## 3, 5, "under_level" set;
2. 12,4 – detectors ##1, 2, 3, 5, "under_level" set;
3. 11,5 – total sum of detectors ## 1, 2, 3, and 5.

The Gaussian contribution does not grow, while the total number of events changes from 43 to 104.

One might argue that the peak seen in 30 percents of statistics and not seen in 70 percents of it is just a trick of statistics. Evaluation shows that probability of this case is less than 1 percent. We are forced to conclude, that appearance of this peak does not correspond to any decay line in this energy range and is not connected with statistics, and cannot be considered as any evidence of neutrinoless double beta-decay.

## Conclusion

### 2β2ν - mode

Our result for the half-life of 2β2ν - mode is as follows

$$T_{1/2}(2\beta 2\nu) = (1.78 \pm 0.01\,[\text{stat}]\,^{+0,07}_{-0,09}\,[\text{sist\_1}] \pm 0.01\,[\text{sist\_2}]) * 10^{21}\,y\quad(68\%\,\text{C.L.}),\quad(4)$$

where the difference between two samples, presented in figure 6 is included as an additional systematic error [sist_2].

### 2β0ν - mode

Concerning neutrinoless mode it is necessary to stress the following:
- an "evidence" of 2039 keV peak could be seen the 3d and 5th detectors' data and only in the runs with under threshold pulses;

- all other runs of 4 assembled detectors (70% of statistics) do not give any evidence of 2039 keV peak. That is seen from figure 7 and confirmed from analysis with Bayesian algorithm.

To estimate half-life for neutrinoless mode, we have used all data in 2000 –2080 keV window. Total energy resolution in 2039 keV range is (4.23 ± 0.14) keV, the background (without peaks of Bi-214) is 0.163 counts/kg*year*keV. Number of events, expected in the interval of ± 3σ is (87 ± 3), quantity of events measured in experimental spectrum is 89. Using the conservative procedure of estimation, as recommended in [3], we will get

$$T_{1/2}(2\beta 0\nu) \geq 1,55 * 10^{25} \text{ y} \quad (90\% \text{ C.L.}) \qquad (5)$$

Consideration of the analysis of pulse shape can give additional opportunities to analyze experimental spectra. Since method of pulse shape analysis with the help of neuron network has not been completely worked out yet, we do not perform the results of this analysis. Detailed comparative analysis for different pulse shape analysis methods and estimation of their sensitivity is performed in [4].

**Acknowledgment**

The work was presented at NANP-2003, Dubna
We are pleasure to thank all participants for useful discussions, especially Danilov M.V, Zeldovich O.Ya. and Kirpichnikov I.V..